\documentclass[superscriptaddress,showpacs,aps,twocolumn,prb,floatfix]{revtex4-1}
\usepackage{amssymb}
\usepackage{amsmath}
\usepackage[dvips]{graphicx}
\usepackage{bm}

\begin{document}
\title{Relation between width of the zero-bias anomaly and Kondo temperaure in 
transport measurements through correlated quantum dots: Effect of asymmetric coupling to the leads}

\author{D. P\'erez Daroca}
\affiliation{Gerencia de Investigaci\'on y Aplicaciones, Comisi\'on Nacional de
Energ\'ia At\'omica, (1650) San Mart\'{\i}n, Buenos Aires, Argentina}
\affiliation{Consejo Nacional de Investigaciones Cient\'{\i}ficas y T\'ecnicas,
(1025) CABA, Argentina}

\author{P. Roura-Bas}
\affiliation{Centro At\'{o}mico Bariloche, Comisi\'{o}n Nacional
de Energ\'{\i}a At\'{o}mica, 8400 Bariloche, Argentina}
\affiliation{Consejo Nacional de Investigaciones Cient\'{\i}ficas y T\'ecnicas,
(1025) CABA, Argentina}

\author{A. A. Aligia}
\affiliation{Centro At\'{o}mico Bariloche, Comisi\'{o}n Nacional
de Energ\'{\i}a At\'{o}mica, 8400 Bariloche, Argentina}
\affiliation{Instituto Balseiro, Comisi\'{o}n Nacional
de Energ\'{\i}a At\'{o}mica, 8400 Bariloche, Argentina}
\affiliation{Consejo Nacional de Investigaciones Cient\'{\i}ficas y T\'ecnicas,
(1025) CABA, Argentina}

\begin{abstract}
The zero-bias anomaly at low temperatures, originated by the Kondo effect when an electric current flows through 
a system formed by a spin-$1/2$ quantum dot and two metallic contacts is theoretically investigated.
In particular, we compare the width of this anomaly $2T_{\rm NE}$ with that of the Kondo resonance in the 
spectral density of states $2T_{K}^{\rho}$, obtained from a Fano fit of the corresponding curves and also with 
the Kondo temperature, $T_K^G$, defined from the temperature evolution of the equilibrium conductance $G(T)$.
In contrast to $T_K^G$ and $2T_{K}^{\rho}$, we found that the scale $2T_{\rm NE}$ strongly depends on the 
asymmetry between the couplings of the quantum dot to the leads while the total hybridization is kept constant.
While the three scales are of the same order of magnitude, $2T_{\rm NE}$ and $T_{K}^{\rho}$ agree only in the case 
of large asymmetry between the different tunneling couplings of the contacts and the quantum dot. 
On the other hand, for similar couplings, $T_{\rm NE}$ becomes larger than $T_{K}^{\rho}$, reaching the maximum 
deviation, of the order of $30\%$, for identical couplings.
The fact that an additional parameter to $T_{\rm NE}$ is needed to characterize the Kondo effect,
weakenig the universality properties, 
points that some caution should be taken in the usual identification in experiments
of the low temperature width of the zero-bias anomaly with the Kondo scale. 
Furthermore, our results indicate that the ratios $T_{\rm NE}/T_K^G$ and $T_{K}^{\rho}/T_K^G$ depend on the 
range used for the fitting.
\end{abstract}


\maketitle

\section{Introduction}

The Kondo effect is one of the most relevant examples of the nontrivial role
of correlations in quantum many-body systems.\cite{kondo,hewson-book}
Initially observed in magnetic impurities embedded in metals, is nowadays
the most interesting regime and often found at low temperatures when
measuring the electric current through quantum dots (QDs) in semiconducting
materials,\cite{gold,cronenwett,gold2,wiel,grobis,kreti,ama,keller} carbon
nanotubes,\cite{jari,choi1,lim,ander,lipi,buss,fcm,grove} and molecular
systems,\cite{liang,yu,leuen,parks,roch,scott,parks2,serge,vincent,yelin} in
which the QD acts as the magnetic impurity. In their more usual
realizations, the Kondo effect can be understood as the screening of the
impurity magnetic moment by the surrounding free conduction electrons
forming a many-body singlet. A remarkable feature of this phenomena is given
by its universality. Different physical properties depending on temperature, 
$T$, bias voltage, $V_b$, and magnetic field, $B$, among others, display an
universal behavior once they are properly scaled by the Kondo temperature $T_K$.\cite{grobis,kreti}

The Kondo temperature is a many-body energy scale (here we take the
Boltzmann $k_B=1$) that can be thought as the binding energy of the spin
singlet. A precise determination of this energy is always desirable. For the
simplest theoretical case in which a single interacting spin degenerate
level, at energy $E_d$ below the Fermi one, is coupled via the hopping, $V$,
to the conduction electrons there is a well defined analytical expression of
this magnitude.\cite{hewson-book} The same is true for two-level or two-dot
generalizations with SU(4) symmetry like QDs in carbon nanotubes,\cite{jari,choi1,lim,ander,lipi,buss,fcm,grove} or silicon nanowires,\cite{tetta}
or a system with two capacitively coupled QDs.\cite{keller,restor,nishi}
Instead, for a QD with several levels coupled with two single-band leads,
analytical expressions for $T_K$ are more difficult to obtain and when more
than one level is occupied a scenario with several stages of the Kondo
effect is the general situation.\cite{pust,desint,karki} In addition, the
Kondo temperature scale can be obtained numerically from the low-temperature
behavior of several magnitudes, like thermodynamics properties, for instance
the impurity entropy,\cite{dinapoli-prl,dinapoli-prb} or dynamical ones,
like the width of the Kondo peak (the one located near the Fermi energy) in
the impurity spectral density.\cite{tosi-orbital-kondo} The resulting $T_K$
obtained from different properties are different, although of the same order
of magnitude.

In general, and particularly for the one-level single-channel system,
measurements of the electrical current through the impurity, $J$, and its
derivative with respect to the bias voltage $V_b$, the conductance $G(V_b,
T)=dJ/dV_b$, characterize the Kondo phenomenon. At low enough temperature, 
$G(V_b, T)$ as a function of the bias voltage has a peak at $V_b = 0$, the
zero-bias anomaly (ZBA). The peak has a width which is narrow compared to
the other energy scales of the system. Increasing $T$ broadens the peak
until it completely disappears. In addition, under an applied magnetic
field, a splitting of ZBA appears. These properties of the ZBA represent the
most clear evidence of the Kondo effect.\cite{cronenwett}

In case of spin-$1/2$ QDs, the Kondo temperature is commonly extracted from
i) fitting the temperature-dependence of the equilibrium conductance 
$G(0,T)=dJ/dV_b \big\vert_{V_b = 0}$, which follows a phenomenological
expression obtained using the numerical renormalization group \cite{gold,G_E}

\begin{equation}
G(0,T)=\frac{G_{s}}{\left[ 1+\left( 2^{1/s}-1\right) \left(
T/T_{K}^{G}\right) ^{2}\right] ^{s}},  \label{tk-from-eq}
\end{equation}
where $s=0.22$, and $G_{s}$ is the conductance at temperature $T=0$ being 
$T_{K}^{G}$ the only adjustable parameter and, ii) extracted from a fitting
of the full width at half maximum (FWHM) of the zero-bias anomaly using the
expression \cite{tsukahara,girovsky} (using the electron charge $|e|=1$)

\begin{equation}
\mathrm{FWHM}(T)=\sqrt{(\beta T)^{2}+(2T_{K})^{2}},  \label{tk-from-width}
\end{equation}
with $\beta $ an extra fitting parameter. This expression for the FWHM gives
the result $2T_{K}$ at zero temperature. Usually, to define the FWHM, the Fano 
formula \cite{fano} is used, which at very low temperatures directly relates the
width of the ZBA with a Kondo temperature

\begin{equation}
G(V_{b},T=0)=C\frac{(\epsilon +q)^{2}}{1+\epsilon ^{2}},~~\epsilon =\frac{V_{b}-\epsilon _{0}}{\Gamma },  
\label{fano}
\end{equation}
where $\epsilon_{0}$ is the center of the ZBA. The parameter $q$ represents the
degree of asymmetry in the line shape which continuously evolves from a dip
for $q=0$ to a peak for $q\rightarrow +\infty $ \cite{Ujsaghy}. In both
limiting cases, the fitting function reduces to a constant plus a Lorentzian,  
and $\Gamma $ is the half width of the dip or peak. Usually in experiments $\Gamma$ 
is identified as the Kondo scale, which we denote as $T_{\mathrm{NE}}$. 

A similar Fano fit can be done for the spectral density of the impurity $\rho (\omega )$, 
leading to a third possible definition of the Kondo scale $T_{K}^{\rho }$. 

One might argue that since the Kondo effect is an equilibrium phenomenon, which is destroyed
by an application of a large bias voltage $V_b$,  a quantity such as  $T_{\mathrm{NE}}$
obtained from non-equilibrium measurements is not a good representation of 
the Kondo scale. However, calculations using perturbative renormalization group 
and poor man’s scaling, valid when the largest of $eV_b$ and the Zeeman energy $B$
is much larger that the equilibrium Kondo scale $T_K$, find that $G(V_b)$ 
and the magnetization are universal functions of $eV_b/T_K$ and 
$B/T_K$.\cite{rosch} In Ref. \onlinecite{plety}, the authors use using real-time
renormalization group calculations to propose a scaling function 
for $G(V_b)$ more involved than Eq, (\ref{fano}).
Therefore in principle $T_K$ can be extracted from non-equilibrium measurements.
We note that both works assumed symmetric coupling to the leads.

The main message of our work is that while $T_{K}^{G}$ and $T_{K}^{\rho }$
do not depend on the asymmetry of the coupling of the QD to the leads, the
width of the zero-bias anomaly $2T_{\mathrm{NE}}$ does. Therefore,
part of the universality is lost, since for example the dependence of
different quantities on magnetic field or temperature, do not depend only on 
$T_{\mathrm{NE}}$ but also on the asymmetry ratio. Then, $T_{\mathrm{NE}}$
cannot be considered as a Kondo temperature, although it is closely related
to this concept. We also show that the result of the Fano fit Eq. (\ref{fano}) 
depends markedly on the range of values chosen for the fit. Both results
are relevant for experiments, as discussed below.

The half width at half maximum of $\rho (\omega )$ is other frequently used
definition of the Kondo temperature, but this quantity is difficult to
access experimentally. Nevertheless, the spectral density has been measured
in a three-terminal quantum ring,\cite{letu} and a splitting of Kondo
resonance for a high enough bias voltage has been observed. A problem of
using the half width at half maximum is that it depends on the subtraction
of a background. We discuss this point in more detail in Section \ref{fitpro}. 
Another possibility to define a Kondo scale is from the dependence of the
conductance for $V_{b},T,B\rightarrow 0$, where $B$ is the magnetic 
field.\cite{grobis,scott,ogu2,rinc,rati,sela,roura,sca,mu,cb,ct} We would like to
mention here that while expanding Eq. (\ref{tk-from-eq}) in powers of $T$
leads to the correct quadratic dependence of the deviation $G(T)-G(0)$, the
coefficient is not correct.\cite{ct} Since the concept of scaling is usually
used for a whole curve and not just a leading behavior, we prefer to use $T_{K}^{G}$ 
as the Kondo scale rather than a similar quantity derived from
some leading term in the expansion of the conductance. For this reason, we
restrict the discussion in our paper to the relation between $T_{\mathrm{NE}}$ 
and the Kondo scales $T_{K}^{G}$ and $T_{K}^{\rho }$.

In Ref. \onlinecite{wiel}, W. G. van der Wiel \textit{et al.}, pointed out
that applying a finite bias voltage introduces dephasing even at very low
temperatures which leads to a possible deviation of $T_{\mathrm{NE}}$ from
the values of $T_{K}^{G}$ obtained from Eq. (\ref{tk-from-eq}). In this
work, we discuss this deviation and show that there is, in addition, a
geometrical mechanism that also introduces differences in the magnitudes
extracted from Eqs. (\ref{tk-from-eq}), (\ref{tk-from-width}) and (\ref{fano}). 
This is the asymmetry between the tunneling couplings of the QD and the
leads. Previous works have studied the relation between zero-bias anomaly
and Kondo temperature, for instance in Refs. \onlinecite{plety,klochan,capac}. 
However, the effect of asymmetry has not been discussed in detail.

We represent the QD by the spin-$1/2$ Anderson impurity model (AIM) and
study the differential conductance by using the non-crossing approximation
(NCA) in its non-equilibrium extension. To complement the results,
particularly at low temperatures, we also use renormalized perturbation
theory (RPT). As discussed in Section \ref{model} these approaches are
complementary. We obtain that the Fano fit depends on the range of values
used in the fitting procedure. This is supported by calculations using the
numerical renormalization group (NRG). We also find out that $T_{\mathrm{NE}}
$ varies with the asymmetry while the total coupling is set to be constant.
This behavior is against the expected universality of the Kondo scale, which
should roughly depend only on the energy level position and the total
coupling. We find that $T_{\mathrm{NE}}$ is approximately 30\% larger than $T_{K}^{\rho }$ 
for symmetric couplings and only tends to same value for
large asymmetry. 

Our results are relevant for spectroscopic measurements using a scanning
tunnel microscopy (STM) performed over magnetic impurities (atoms and
molecules) deposited on metallic surfaces. In these measurements the
symmetry of the couplings between the magnetic impurity and the surface, and
the magnetic impurity and the tip is an important issue. In particular,
experiments where the tip is moved over the surface, are examples of the
change in the symmetry of the aforementioned couplings. For instance, in
Ref. \onlinecite{lorente} such measurements are made for a system consisting
in a Co atom on Cu(111) and the tip of the STM is moved vertically on top of
the Co atom until contact. The authors characterized the point of contact by
the physical situation at which the curve $G(V_{b})$ vs $V_{b}$ becomes
symmetric, which in turn means that both couplings, Co-Cu and Co-tip are
approximately equal.\cite{capac,diego} In the experiment, the width of $%
G(V_{b})$ varies due to the monotonic increase of the coupling Co-tip when
the tip is moved towards the Co atom and also for the more subtle variation
of the asymmetry between Co-Cu and Co-tip couplings, which seems to have
been missed in previous work. A similar experimental procedure is used for a
system consisting on a Co atom adsorbed on Cu(100) and Cu(111) surfaces, as
described in Refs. \onlinecite{choi, choiPRL,neelPRL,neelPRB}.

We find that $T_{\mathrm{NE}}$ deviates from $T_{K}^{\rho }$ in experimental
setups for which the tunneling couplings between QD and the leads are
approximately the same. Furthermore, the relations of both quantities to $T_{K}^{G}$ 
given by Eq. (\ref{tk-from-eq}) depend on the range of voltages
or energies used in the fitting procedure. This is important, since $T_{\mathrm{NE}}$ 
is widely used as an estimation of $T_{K}$ in different
classes of experiments.\cite{perera,choi,schneider,karan,bergmann,karan2,karan3,maria} For example, in
Ref. \onlinecite{schneider} a "Kondo temperature" $T_{\mathrm{NE}}=92$ K for
a Co impurity on Ag(111) is reported, while in Ref. \onlinecite{maria}\ \ a
distribution of \ $T_{\mathrm{NE}}$\ with an average 52.1 $\pm$ 9.4 K was
found for the same system. While part of the discrepancy might be due to the
variation of the surface density of states,\cite{maria} the nearly four times wider range of
voltages used in Ref. \onlinecite{schneider} in the fitting procedure can
explain the different $T_{\mathrm{NE}}$ as we shall show.

The paper is organized as follows. In Sec. \ref{model}, we describe the IAM
and the method we used. In Sec. \ref{results} we discuss the properties of
the differential conductance and its dependence with the couplings to the
leads. Sec. \ref{sum} contains a summary and a discussion.

\section{Model and methods}

\label{model} As we have mentioned, we use the spin-$1/2$ impurity AIM to
describe the molecular or semiconducting QD. This choice does not limit the
qualitative validity our findings. In Section \ref{sum} we address the more
general case of large values of the molecular spin and also total and
partial screening of it.

The model is composed by a single level characterized by an energy $E_{d}$
below the Fermi energy (which we choose at the origin of energies) coupled
to two conduction leads. The Hamiltonian reads as follows 
\begin{eqnarray}
H &=& E_{d}n_{d}+U n_{d\uparrow} n_{d\downarrow }
+\sum_{\nu k\sigma }\epsilon _{k}^{\nu }c_{\nu k\sigma}^{\dagger }c_{\nu k\sigma }  \notag \\
&&+\sum_{\nu k\sigma }(V_{k}^{\nu }d_{\sigma }^{\dagger }c_{\nu k\sigma }+
\mathrm{H.c}.)  \label{ham}
\end{eqnarray}
with $n_{d}=\sum_{\sigma }n_{d\sigma }$, $n_{d\sigma }=d_{\sigma }^{\dagger
}d_{\sigma }$. The operator $d_{\sigma }^{\dagger }$ creates an electron
with spin $\sigma $ at the single level of the QD while the operators 
$c_{\nu k\sigma }^{\dagger }$ create conduction electrons at the leads.
Depending on the specific experiment, they can represent left and right
leads when conduction through a QD is studied, or the metallic substrate ($\nu =S$) 
and the tip ($\nu =T$) of the STM in scanning tunneling
spectroscopy. The parameters $V_{k}^{\nu }$ describe the hopping elements
between the leads and the QD. For most of our results, we take the value of
the Coulomb repulsion to be infinite, $U\rightarrow \infty $ and analyze the
model within the Kondo regime, $-E_{d}\gg \Delta $, being $\Delta $ the
resonant level half-width. Finite values of $U$ within this regime only
change the Kondo scale while the present analysis remains valid.

In the case of having different chemical potentials $\mu_\nu$ in the
metallic contacts, a constant electric current flows through the QD in the
steady state. We take the same temperature $T$ for all elements of the setup
and fix the chemical potentials to be $\mu_\nu=-e\gamma_{\nu}V_b/2$ with the
sign $\gamma_{\nu}=-(+)$ for $S(T)$ being $V_b$ the bias voltage, as a
reference.

The charge current through the QD is given by \cite{meir, win, ang} 
\begin{eqnarray}  \label{currents}
J&=& \frac{2e\pi}{h}A(\alpha)\Delta \int d\omega \rho (\omega )\left(
f_{S}(\omega)-f_{T}(\omega)\right).
\end{eqnarray}
Here the energy $\Delta$ incorporates both, the substrate-dot and tip-dot
couplings, $\Delta=\Delta_S + \Delta_T$, being $\Delta _{\nu }=\pi
\sum_{k}|V_{k}^{\nu }|^{2}\delta (\omega -\epsilon _{k}^{\nu })=\pi V_{\nu
}^{2}\rho_{\nu} $ assumed independent of energy. 
Furthermore, $f_{\nu}(\omega)=1/[\exp({\omega-\mu_{\nu}}/{T}) + 1)]$ is the
Fermi distribution associated to the lead $\nu$, and the spectral function
of the QD per spin, is denoted by $\rho (\omega )$. Regarding the factor 
$A(\alpha)=4\alpha/(\alpha+1)^2$, it represents the asymmetry in the device
geometry being $\alpha=\Delta_S / \Delta_T$ the ratio of the tunneling
couplings.

For the calculation of $\rho (\omega )$ entering Eq. (\ref{currents}), we
mainly use the non-equilibrium non-crossing approximation (NCA) \cite{win,roura_1,tosib}. 
The non-equilibrium NCA technique is one of the
standard tools for calculating the spectral density of the dot within the
Kondo regime in which the population of the dot is near 1. NCA has being
successfully applied to the study of a variety of systems such as two-level
QD's and C$_{60}$ molecules displaying a quantum phase transition,\cite{serge,tosi-orbital-kondo,roura_2,roura2010} 
or a nanoscale Si transistor \cite{tetta} among
others. Few alternatives exist out of equilibrium, like renormalized
perturbation theory \cite{ang,hbo,ogu,cb,ct}, Fermi-liquid approaches \cite{fili} 
and slave bosons,\cite{sierra,sierra2} which are restricted to small
voltage and temperature, equations of motion with some difficulties to
reproduce the Kondo energy scale,\cite{rapha,rome2,li} or real-time
renormalization group.\cite{plety,klochan} Recently, a variational approach has been proposed.\cite{ashi}

Nevertheless, the NCA has some
limitations at very low temperatures (typically below $0.05T_{K}^{G}$). In
particular, it does not satisfy accurately the Friedel sum rule at zero
temperature.\cite{fcm} For this reason, we also used the approach of
renormalized perturbation theory (RPT) used before by one of us.\cite{ang,ct}
It consists of using renormalized parameters for $E_{d}$, $U$ and $\Delta $
obtained at $V_{b}=T=0$ by a numerical-renormalization-group calculation,\cite{cb,note} 
and incorporating perturbations up to second order in the
renormalized $U$. At equilibrium, the method provides results that coincide
with state-of-the art techniques for the dependence of the conductance with
magnetic field $B$ ($c_{B}$) \cite{cb} and temperature ($c_{T}$) \cite{ct}
to second order in $B$ or $T$. An analytical expression for $c_{T}$ in terms
of the renormalized parameters was provided.\cite{ct} However, for energy
scales of the order of $T_{K}^{G}$ or larger, the method loses accuracy and
in particular if fails to give a splitting of the spectral density for 
$eV_{b}>T_{K}^{G}$, which is however well reproduced by the NCA.

\section{Results}

\label{results}

In what follows we set $\Delta =1$ as our unity of energy,  $E_{d}=-4$ for
the energy level of the QD and $U\rightarrow \infty $, unless otherwise
stated. Some results are presented with $E_{d}=-6$, and some RPT results in
the symmetric case $U=8$, $E_{d}=-4$ are also shown. The choice of $E_{d}=-4$
does not affect our discussion as long as the Kondo regime $|E_{d}|\gg
\Delta $. As usual, we assume a constant conduction density of states with
bandwidth $2D$. We use $D=10$.

\subsection{Nonequilibrium conductance}

\label{nec}

We start our discussion by giving a brief description of the experiment
recently made by Choi \textit{et al}. in Ref. \onlinecite{choi}, in which a
Co atom is deposited on a Cu(111) surface. A STM with a tip that also
contains Cu is placed vertically over the adsorbed Co atom and is used to
measure the tunneling current. From a distance tip-Co large enough,
characterized by a tip-Co coupling $\Delta_T\ll\Delta_S$, the authors move
the tip towards the surface till contact, which is defined by the condition
of getting a symmetric curve of the conductance as a function of bias
voltage $G(-V_b)=G(V_b)$.

\begin{figure}[tbp]
\includegraphics[clip,width=8cm]{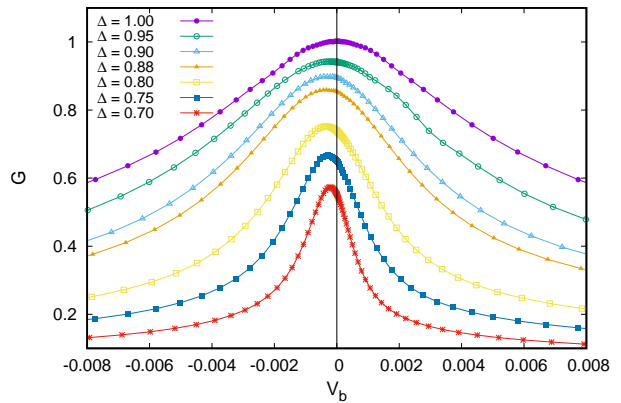}
\caption{(Color online) Differential conductance as a function of the
applied bias voltage for $\Delta_S = 0.5$ and several values of $\Delta_T$
from $0.2$ (largest distance) to $0.5$ (point contact).}
\label{GammaR_varia.eps}
\end{figure}

In Fig. \ref{GammaR_varia.eps} we show the differential conductance $G(V_b,T)
$ as a function of bias voltage $V_b$ for several values of $\Delta_T$ and
temperature $T$ low enough so that the conductance has already reached the
saturated value for $T=0$ (in practice we have taken $T \sim T_K^G /20$
where $T_K^G$ depends on $\Delta_T$). As expected, as $\Delta_T$ increases, 
$G$ also increases and becomes more symmetric, in qualitative agreement the
results already presented in Fig. 1(c) of Ref. \onlinecite{choi}. 
The different curves, from bottom to top, represent the excursion of the tip
as vertically approaches the adsorbed Co atom. We have fixed the
hybridization Co-Cu to be $\Delta_S = 0.5$. We simulate larger distances
between the Co atom and the STM tip by smaller coupling $\Delta_T$. As soon
as the distance is reduced, the value of $\Delta_T$ increases until the
point contact is reached. We define the point contact by $\Delta_T=\Delta_S$
giving a total coupling of $\Delta = 1$. 

The main features of the different curves are that the width, intensity and
symmetry increase as $\Delta$ does. The increase of the width is related to
the increase of the Kondo scale, which in turn, depends on the total
coupling $\Delta$. The increase of the intensity is due to the increase of
the asymmetry factor $A(\alpha)$, which reaches $A(\alpha)=1$ for the point
contact. Finally, the symmetry increases as $\Delta_T$ does. This is related
with the fact that for $\Delta_T\ll\Delta_S$, $G(V_b)$ mimics the spectral
density $\rho(\omega)$ which is in turn asymmetric due to the infinite value
of $U$,\cite{capac,diego} while for the opposite limit, $\Delta_T\sim\Delta_S
$, $G(-V_b)=(V_b)$ as a consequence the reflection S $\leftrightarrow$ T
symmetry.

\begin{figure}[tbp]
\includegraphics[clip,width=8cm]{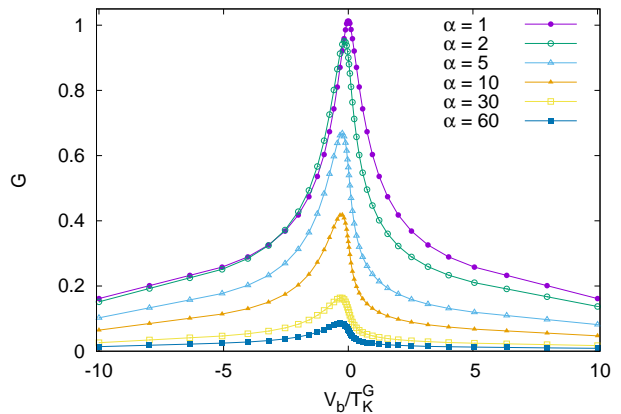}
\caption{(Color online) Differential conductance as a function of the
applied bias voltage for $\Delta = 1$ and several values of $\protect\alpha$.
}
\label{dIdV_Tmin_alfas.eps}
\end{figure}

The increase in the width of the differential conductance as $\Delta _{T}$
increases is due to the addition of two effects that cannot be disentangled
in the figure: the increase in $T_{K}^{G}$ and the decrease in the asymmetry
of the couplings. In order to separate these effects, we show in Fig. \ref{dIdV_Tmin_alfas.eps} 
the results for the differential conductance for a
fixed total coupling $\Delta =1$ as a function of the applied voltage and
the asymmetry ratio $\alpha =\Delta _{S}/\Delta _{T}$. In this way, the
results become independent of $\Delta $ but retain the dependence in the
asymmetry, which is the main ingredient in the present discussion. As in
Fig. \ref{GammaR_varia.eps}, the symmetry and intensity increase as $\alpha $
is reduced. However, it is expected that the Kondo temperature remains
constant simply because it depends on the total coupling $\Delta $ and not
on the asymmetry ratio $\alpha =\Delta _{S}/\Delta _{T}$. From the spectral
density, the Kondo temperature is given by the half-width al half-maximum of
the Kondo peak. It is well known that the shape of the Kondo resonance in
the spectral density does not depend on the asymmetry coupling, see for
instance Fig. 1 in Ref. \onlinecite{diego}. We have verified that this is
actually the case for the whole values of $\alpha $ from $\alpha =1$ to $\alpha =60$. 
Therefore, the width of this resonance, or that obtained from
the Fano fit, which we denote as $2T_{K}^{\rho }$, is independent of $\alpha 
$. However from the figure, particularly for small values of $\alpha $, it
is clear that the width of the peak in the differential conductance $G(V_{b})
$ ($2T_{\mathrm{NE}}$ from a Fano fit) \textit{narrows} as $\alpha $
increases.

\begin{figure}[tbp]
\includegraphics[clip,width=8cm]{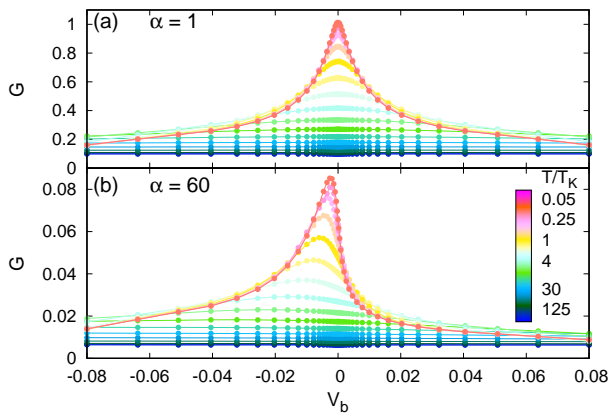}
\caption{(Color online) Temperature evolution of the differential
conductance as a function of the applied bias voltage for two selected
values of the asymmetry parameter $\protect\alpha$. The temperature range
covers the region from $125T_K^G$ to $T_K^G/20$.}
\label{dIdV_Ts.eps}
\end{figure}

\begin{figure}[tbp]
\includegraphics[clip,width=8cm]{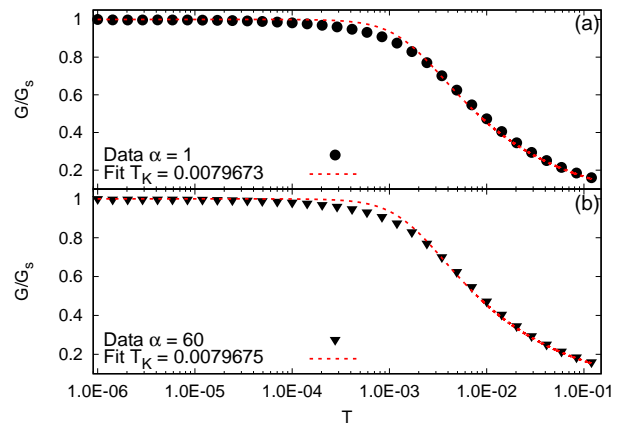}
\caption{(Color online) Temperature evolution of the equilibrium conductance
(discrete points) normalized by its saturation values and the corresponding
fitting with Eq.\protect\ref{tk-from-eq} (full red lines).}
\label{ajuste.eps}
\end{figure}

The temperature evolution of the differential conductance can also be used
to determine the Kondo scale $T_K^G$ by using Eq. (\ref{tk-from-eq}). In
Fig. \ref{dIdV_Ts.eps} we show the temperature dependence of $G(V_b)$ for
the symmetric case $\alpha=1$ (top panel), which can be related with the
point contact regime of the experiment in Ref. \onlinecite{choi}, and for
the opposite strong asymmetric one $\alpha=60$. From the figure, it is not
obvious that a fitting of the values of $G(V_b=0)$ (obtained from the
non-equilibrium calculation) with Eq. (\ref{tk-from-eq}) gives the same
result of $T_K^G$ for both cases. However in Fig. \ref{ajuste.eps} we
confirm that this is actually the case. We have verified that the same value
of $T_K^G=0.00797$ is obtained independently of the value of $\alpha$. Fig. \ref{ajuste.eps} 
shows the temperature dependence of the equilibrium
conductance for the two selected values of the asymmetry parameter, $\alpha=1$ and $\alpha=60$. 
Calculating $G(T)$ at $V_b=0$ from an equilibrium calculation gives the same
result.

\subsection{Fitting procedure to determine the widths of the curves}

\label{fitpro}

For a quantitative analysis of the effect of the asymmetry in the widths of
the Kondo resonances in the differential conductance $G(V_b)$ and the
spectral density $\rho(\omega)$ at small temperature, we need some procedure
to determine these widths. Experimentally Fano fits described by Eq. (\ref{fano}) 
after subtracting a background, are the most widely used. The Frota
function \cite{frota} is also used but it does not permit asymmetric shapes
and therefore it is not useful for our purpose. Theoretically, the
half-width at half maximum of the corresponding curve is also used. However,
we have found that this leads to an overestimation of the widths due to the
fact that the Kondo peak is mounted on the tails of the charge transfer
peaks. These peaks in the spectral density are centered at energies $E_d$
and $E_d+U$ and have total width near $4 \Delta$.\cite{capac,anchos}
Therefore, we analyze the widths using the Fano formula rewritten in the
following form using $q=1/x$ in Eq.(\ref{fano}) and adding a constant
background $A$:

\begin{equation}  \label{fano2}
G=A + B\frac{(1+x\epsilon)^2}{1+\epsilon^2},
\end{equation}

\begin{figure}[h]
\includegraphics[clip,width=8cm]{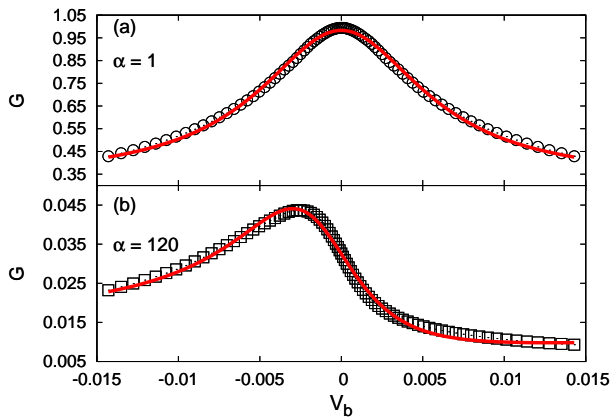}
\caption{(Color online) Differential conductance (discrete points) as a
function of the bias voltage for $\protect\alpha=1$ (top panel) and 
$\protect\alpha=120$ (bottom panel). The continuous line corresponds to the Fano
fitting with Eq. (\protect\ref{fano2}).}
\label{Fano1-120.eps}
\end{figure}

Note that for the symmetric case $\alpha=1$, which corresponds to $x=0$, $G$
is a constant plus a Lorentzian function with half-width $\Gamma$. Examples
of fits of $G(V_b)$ are given in Fig. \ref{Fano1-120.eps}

We identify the value of $\Gamma $ that results from the fit of $G(V_{b})$
at small enough temperatures with $T_{\mathrm{NE}}$. Similarly, the value of 
$\Gamma $ obtained fitting $\rho (\omega )$ gives $T_{K}^{\rho }$. While the
result of the fit is unambiguous, a difficulty of this procedure (also found
experimentally \cite{moro}) is that the resulting $\Gamma $ depends on the
window (range of values of the abscissa) of the fit. In some experimental
work with scanning tunneling microscopy,\cite{moro} the fitting range of $G(V_{b})$, 
$-W\leq eV_{b}\leq W$ was established as $W=1.5T_{\mathrm{NE}}$
obtaining $T_{\mathrm{NE}}$ from the fit and then changing $W$ if it does
not coincide with $1.5T_{\mathrm{NE}}$ until convergence. In our case, it is
simpler to define the range $W$ in terms of the Kondo temperature determined
from the conductance at equilibrium $T_{K}^{G}$, which is defined
unambiguously. 

\begin{figure}[h]
\includegraphics[clip,width=8cm]{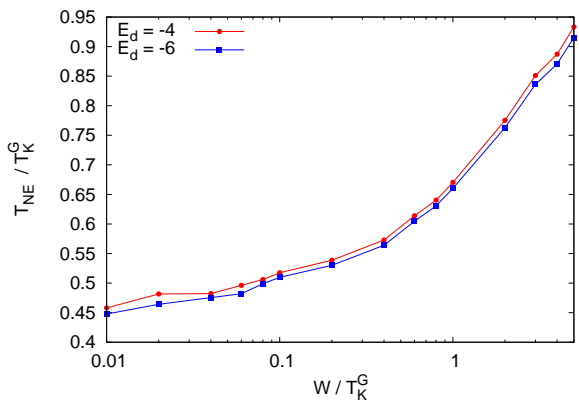}
\caption{(Color online) $T_{\mathrm{NE}}$ obtained from the Fano fit of
the differential conductance as a function of the window of the fit $-W \leq
eV_b \leq W$ for $\protect\alpha=1$ and two values of $E_d$.}
\label{windowv}
\end{figure}

In Fig. \ref{windowv} we show the half-width $\Gamma $ of the ZBA in the
differential conductance obtained from the fit Eq. (\ref{fano2}) for
different fitting windows measured in units of $T_{K}^{G}$. Several
conclusions can be drawn from this figure. In spite of taking the symmetric
case $\alpha =1$, for which $x=0$, the shape of the curve (corresponding to
the top panel in Fig. \ref{Fano1-120.eps}) is not Lorentzian. Otherwise its
width would be independent of the window of the fit. However, normalizing
the width with $T_{K}^{G}$ its shape is universal. It is the same for
different values of $E_{d}$. The same is true for the values of $A$ and $B$
of the fit using Eq. (\ref{fano2}) (not shown). The fit for $W\ll T_{K}^{G}$
becomes meaningless since $A$ tends to the quantum of conductance $G_{0}=2e^{2}/h$ 
and $B$ becomes very small. For $W\gg T_{K}^{G}$ the fit has
too much weight on the tails of the Kondo resonance. The choice $W\sim 1.5T_{\mathrm{NE}}$ 
made by experimentalists \cite{moro} seems reasonable. This
corresponds approximately to $W=2T_{K}^{G}$. The latter choice allows us to
avoid a self-consistent procedure to determine $W$. 

\begin{figure}[h]
\includegraphics[clip,width=8cm]{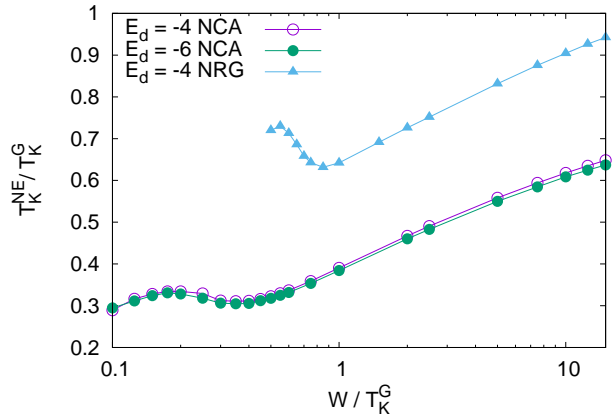}
\caption{(Color online) $T_{K}^{\protect\rho}$ obtained from the Fano fit of
the spectral density as a function of the window of the fit $-W \leq eV_b
\leq W$ for two values of $E_d$ and two techniques.}
\label{windowr}
\end{figure}

Motivated by the arguments above, we take $W=2T_{K}^{G}$ 
for all the calculations of $T_{\mathrm{NE}}$ and $T_{K}^{\rho}$ 
that follow, except in the discussion of Fig. \ref{windowr} 
of the next paragraph.

Qualitatively, the same general features are shared in the fits of the
spectral density $\rho(\omega)$, as it is shown in Fig. \ref{windowr}. We
also show in the figure the results obtained using the numerical
renormalization group (NRG).\cite{nrg} In spite of the known different shape
between the spectral densities calculated by NCA and NRG,\cite{compa} one
can see that qualitatively the same trend of increasing $T_{K}^{\rho}$ with
increasing $W$ for $W \sim T_K^G$ takes place for both approaches. The
difference is that the ratio $T_{K}^{\rho}/T_K^G$ is about 1.5 times greater
with NRG.

\subsection{Comparison of the width of the zero-bias anomaly with the Kondo
scale}

\label{compa}

\begin{figure}[tbp]
\includegraphics[clip,width=8cm]{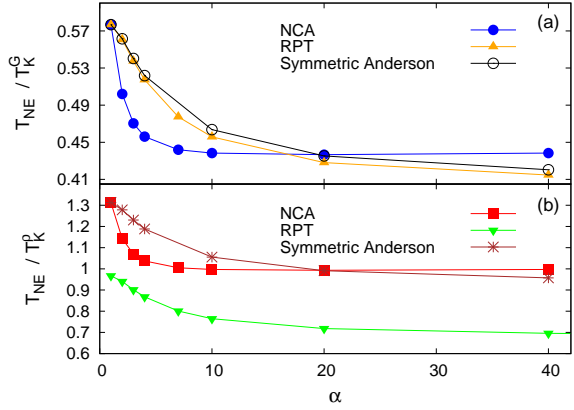}
\caption{(Color online) Ratio of $T_{\mathrm{NE}}$ obtained from the
Fano fit [$\Gamma$ in Eq. (\protect\ref{fano2})] $G(V_b)$ to $T_K^G$ (top
panel) and ratio $T_{\mathrm{NE}}/T_{K}^{\protect\rho}$ (bottom panel)
as a function of $\protect\alpha$ for $T=0.001T_K^G$. Triangles are RPT
results (normalized to the same value for $\protect\alpha=1$ in the top
panel).}
\label{fanos}
\end{figure}

In Fig. \ref{fanos} we show the half width $T_{\mathrm{NE}}$ obtained from
the fit of the differential conductance using the procedure described in
Section \ref{fitpro} as a function of the asymmetry parameter $\alpha $ for
two techniques, NCA and RPT. We discuss first the NCA results. We remind the
reader that for NCA, $T_{K}^{G}=0.00797$ independent of $\alpha $. The
results allows us to quantify the narrowing of $G(V_{b})$ with increasing $\alpha $ 
already apparent in Fig. \ref{dIdV_Tmin_alfas.eps}. The behavior of 
$T_{\mathrm{NE}}$ for moderate and small values of $\alpha $ is unexpected,
and missed in previous studies.\cite{lorente} In the bottom panel of the
figure we show the ratio of the widths derived from the fits of the
differential conductance $G(V_{b})$ and the spectral density $\rho (\omega )$. 
As expected, this ratio tends to 1 for large $\alpha $ and small
temperatures in the Kondo limit. For large asymmetry the dot is practically
at equilibrium with the substrate, so that there is a Kondo peak near the
chemical potential of the substrate $\mu _{S}=eV_{b}/2$, and $G(V_{b})$
mimics the spectral density $G(V_{b})\sim \frac{e^{2}}{h}\pi \Delta A(\alpha
)\rho (\mu _{T})$ with $\mu _{T}=-eV_{b}/2$ the chemical potential of the
tip.\cite{diego} Instead, in the case of symmetric couplings, like the case
of point contact in the experiment of Ref. \onlinecite{choi}, we obtain that
the width of the differential conductance at low temperatures $2T_{\mathrm{NE}}$ 
is nearly $30\%$ larger than the width $2T_{K}^{\rho }$ of the spectral
density of states. This agrees with previous estimates based on the
half-width at half maximum of the corresponding curves.\cite{capac}

For RPT, we have obtained $T_{K}^{G}$ by a fit with Eq. (\ref{tk-from-eq})
in the range $0<T/\widetilde{\Delta }<0.1$, where $\widetilde{\Delta }$ is the renormalized value
of $\Delta $.\cite{ct}. For larger temperatures the RPT results lie above
the universal curve and are not quantitatively reliable. We have used here
two sets of parameters:  $E_{d}=-4\Delta $ and $U\rightarrow \infty $  as
in the NCA calculation [the corresponding RPT parameters are $\widetilde{\Delta }=0.00579$, 
$\widetilde{E}_{d}/\widetilde{\Delta }=0.161$ and $\widetilde{U}/(\pi \widetilde{\Delta })=1.025$ 
(Ref. \onlinecite{ct})], 
and $E_{d}=-4\Delta $, $U=8$ corresponding to the symmetric Anderson model [with
renormalized parameters $\widetilde{\Delta }=0.120$, 
$\widetilde{E}_{d}/\widetilde{\Delta }=0$ and $\widetilde{U}/(\pi \widetilde{\Delta })=0.985$ 
(Ref. \onlinecite{ct})]. From the fitting procedure we obtain 
$T_{K}^{G}=0.7612\widetilde{\Delta }=0.00441$ for $U\rightarrow \infty $ and 
$T_{K}^{G}=0.7390\widetilde{\Delta }=0.0887$ for $U=8$. By comparison the Bethe ansatz expression of the
Kondo temperature for this case (see for example Eq. (8) of Ref. \onlinecite{li})
gives $T_{K}=0.105.$ The dependence with $\alpha $ is qualitatively similar
with that obtained by NCA, particularly for $U=8$. However, for $U\rightarrow \infty $, 
the ratio $T_{\mathrm{NE}}/T_{K}^{\rho }$ is smaller
and in particular it is smaller than 1 in the limit of large $\alpha $ which
indicates a failure of RPT. In the symmetric case $U=-2E_{d}$, the real part
of the renormalized retarded self-energy vanishes by symmetry, while for any parameters,
the renormalized lesser and greater self-energies and the imaginary
part of the renormalized retarded self-energy are exact to total second
order in $\omega $, $T$ and $V_{b}$.\cite{ang} Therefore one expects that
RPT is more accurate in the symmetric case. The symmetric point is
important because often experiments on quantum dots are tuned to this point
at which the spectral density gets its maximum value $1/(\pi \Delta )$
according to the Friedel sum rule,\cite{ang} leading at $V_{b}=T=0$ to 
$G=dI/dV=(2e/h)A(\alpha )$ in Eq. (5). 
From here, the asymmetry factor $A(\alpha )$ can be deduced. 

\begin{figure}[tbp]
\includegraphics[clip,width=8cm]{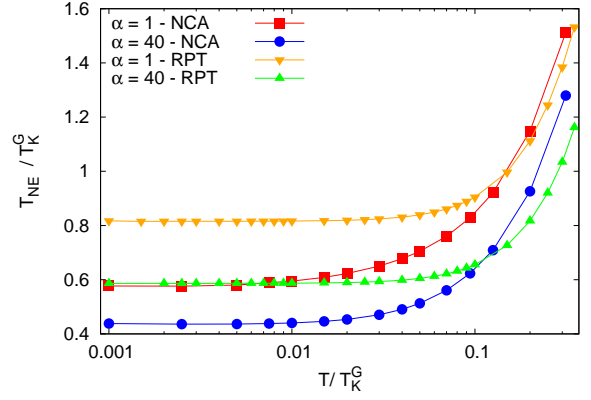}
\caption{(Color online) Ratio $T_{\mathrm{NE}}/T_K^G$ as a function of
temperature for two values of $\protect\alpha$ and two techniques.}
\label{temp}
\end{figure}

In Fig. \ref{temp} we show the change with temperature of the half with of
the differential conductance $T_{\mathrm{NE}}$ as a function of voltage
obtained from the Fano fit Eq. (\ref{fano2}) for NCA and RPT for $U\rightarrow \infty $. 
There is a strong temperature dependence. $T_{\mathrm{NE}}$ increases by a factor near 3 
when the temperature reaches values of
the order of $T_{K}^{G}$. This should be taken into account when Fano fits
are performed on experimental data at finite temperatures. As expected from
Fermi liquid theories,\cite{ang,hbo,ogu,ogu,cb,ct,fili} the dependence
resulting form RPT is quadratic for small temperature. A fit for the data
for $T/T_{K}^{G}<0.1$ gives 
$T_{\mathrm{NE}}(T)/T_{\mathrm{NE}}(0)=1+11.0(T/T_{K}^{G})^{2}$ for $\alpha =1$ and 
$T_{\mathrm{NE}}(T)/T_{\mathrm{NE}}(0)=1+12.0(T/T_{K}^{G})^{2}$ for $\alpha =40$. In contrast, the
NCA results in the same temperature range display a dependence more similar
to a linear one, which is likely to be related with the shortcomings of the
NCA at low temperatures.

\begin{figure}[tbp]
\includegraphics[clip,width=8cm]{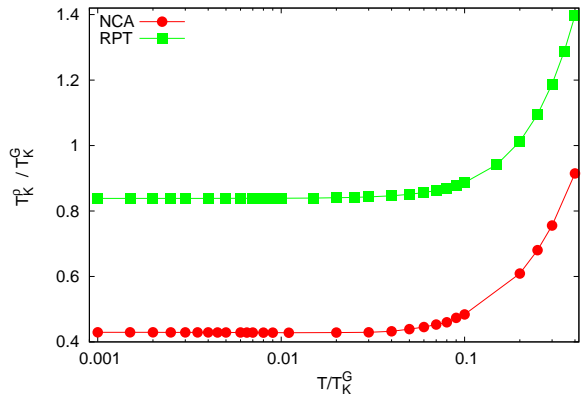}
\caption{(Color online) Ratio $T_{K}^{\protect\rho}/T_K^G$ as a function of
temperature for any $\alpha$ and two techniques.}
\label{temprho}
\end{figure}

For comparison, in Fig. \ref{temprho} we show the temperature dependence of
the half width of the spectral density. As expected, it is weaker than that
of the differential conductance, because in the latter the effects of
broadening of the spectral density and the Fermi functions in Eq. (\ref{currents}) 
are added. Here the low temperature dependence predicted by RPT
for $T/T_K^G < 0.1$ is $T_{K}^{\rho}(T)/T_{K}^{\rho}(0)= 1 + 5.75(T/T_K^G)^2$.

\section{Conclusions}

\label{sum} In summary, we have investigated theoretically the width of the
zero-bias anomaly, originated by the Kondo effect, when a charge current
flows through a system composed by a spin-$1/2$ quantum dot and two metallic
contacts. In particular, we have compared the energy scale represented by
this width $T_{\mathrm{NE}}$ with definitions of the Kondo temperature
obtained from the width of the Kondo resonance in the spectral density at
equilibrium ($T_{K}^{\rho }$) and from the well established temperature
evolution of the equilibrium conductance $G(0,T)$ ($T_{K}^{G}$). Our results
at low enough temperatures show that $T_{\mathrm{NE}}=T_{K}^{\rho }$ only in
cases of large asymmetry between the different tunneling couplings of the
contacts with the quantum dot. On the other hand, if the couplings tend to
be similar, $T_{\mathrm{NE}}$ becomes larger than $T_{K}^{\rho }$.

The ratio $T_{\mathrm{NE}}/T_{K}^{\rho }$ reaches values as
high as 1.30. Following usual experimental procedures, we have determined
the above mentioned widths using a Fano fit of the line shape. Our results
using NCA and also NRG show that the result depends on the range of values
used in the fit. The temperature dependence of these widths is strong and
stronger for the nonequilibrium conductance $G=dI/dV_{b}$ vs $V_{b}$ than
for the spectral density.

As explained in the introduction, our findings are relevant for a wide range
of different experiments. The asymmetry ratio between the tunnel couplings
is directly related with the intensity of the differential conductance at
zero bias from Eq. (\ref{currents}). In particular, in the Kondo limit at
zero temperature $G(0,0)=2A(\alpha )e^{2}/h$. The effects of this asymmetry
in the width of the zero-bias anomaly has been missed previously and is
particularly relevant in experiments in which the Kondo temperature and the
asymmetry ratio are simultaneously changed. On the other hand different
windows used in Fano fit can explain conflicting reports on the width of the
zero-bias anomaly for the same system.

As message to experimentalists, if the precise value of the Kondo temperature matters,
it is more convenient to extract it from the temperature dependence of the
zero-bias conductance ($G(0,T)$) than from the shape of the zero-bias anomaly
as a function of bias voltage ($G(V_b,0)$). If a fit of the latter is done,
the width depends on the range of the fitting (which should therefore be specified) 
and the asymmetry ratio.

While most of our results were calculated assuming infinite on-site repulsion
$U \rightarrow \infty$. calculations in the symmetric case 
$U = - 2E_d$ and Kondo limit $- 2E_d \gg \Delta$ using
renormalized perturbation theory confirm the main conclusions.
We have also assumed a constant density of conduction electrons $\rho_\nu(\omega)$ 
around the Fermi energy for both leads.
Recent experiments obtain an approximate linear dependence of $\Delta$ with the 
applied gate voltage indicating a variation of the density of states.\cite{svila}
We expect that in this case the main result that the ratio 
$T_{\mathrm{NE}}/T_{K}^{\rho }$ varies from near 1.30 for symmetric coupling
to 1 for very asymmetric coupling remains. We also expect that if the 
variation of $\rho_\nu(\omega)$ around the Fermi level is small on the scale of $T_K$,
the results would be very similar as taking the average density in this scale.
As an example, a step of magnitude $\Delta/2$ in $\Delta$ 
[which simulates the onset of the surface band of noble metals at the (111) surface]
at positions $\omega= \pm \Delta/2$ with $\Delta \gg T_K$ changes $T_K$ by 
nearly three orders of magnitude, but the shape of the Kondo peak 
in the spectral density rescaled with
$T_K$ is very similar, indicating that the this peak
is sensitive to the value of $\rho_\nu(\omega)$ near the Fermi energy
and not to its structure for $|\omega|>T_K$ (see Figs. 6 and 7 in Ref. \onlinecite{step}).

Our results can be extended to other systems. For instance, in cases of
impurity spin $S>1/2$, partially screened by one channel of conduction
electrons, there is a similar temperature dependence of the conductance $G(0,T)$ 
as that given by Eq. (\ref{tk-from-eq}) with the difference of
having other values of the parameter $s$. A table of this parameter for
several values of the impurity spin is given in Ref. \onlinecite{parks2}. On
the other hand, in case of a total compensated spin $S=1$ with two
conduction channels, the corresponding expression is given in Ref. \onlinecite{blesio}. 
It would be interesting to study the ratios of the
different Kondo scales in these cases.

\textit{Acknowledgment}. We are indebted to Germ\'an Blesio for the NRG
calculations presented in Fig. \ref{windowr}. A. A. A, was sponsored by PIP
112-201501-00506 of CONICET and PICT 2013-1045 of the ANPCyT.


\begin{thebibliography}{99}
\bibitem{note} The values of the renormalized parameters given in 
Ref. \onlinecite{cb} have a mistake in one entry and in the values of $\Delta_\rho
$ (not used). The correct table is presented in Ref. \onlinecite{ct}.
\end{thebibliography}
\end{document}